# Auger parameter analysis for TiN and AlN thin films via combined in-situ XPS and HAXPES

O.V. Pshyk*, J. Patidar, C. Cancellieri, S. Siol*

*Empa – Swiss Federal Laboratories for Materials Science and Technology, 8600 Dübendorf, Switzerland*

*Corresponding authors:*
*oleksandr.pshyk@empa.ch*
*sebastian.siol@empa.ch*

**Abstract**

Auger parameter analysis provides in-depth information about electronic and chemical bonding properties of TiN and AlN thin films, which are relevant across a wide range of technologies. Meaningful interpretation and analysis of Auger parameter of these materials has been hindered due to, among other reasons, the absence of reliable references. Here we present a comprehensive study of Auger parameters for TiN and AlN thin films using a dual-source lab-based XPS/HAXPES system equipped with Al Kα and Cr Kα x-ray sources. Due to a large spread of excitation x-ray energy, bulk- and surface-sensitive core-level photoelectrons and Auger transitions are probed. This allows us to study a wide range of Auger and core level emission lines of TiN and AlN. These measurements can serve as references for further identification of chemical state changes, oxidation state or any deviations in the local chemical environment in these materials. UHV sample transfer was employed to minimize surface contamination. Additionally, we demonstrate how common procedures such as ambient air exposure and $Ar^+$ sputter-etching influence the Auger parameters, highlighting the importance of surface preparation in spectroscopic analysis.

**Introduction**

TiN and AlN thin films have found a broad range of technological applications due to their excellent physical and functional properties. AlN thin films in wurtzite structure are used in a broad range of piezoelectric applications such as surface acoustic wave (SAW) or film bulk acoustic resonators (FBAR) due to their high acoustic velocity, chemical resistance, thermal stability, large dielectric breakdown field and relatively high piezoelectric coefficient[1]. TiN thin films in face-centered cubic NaCl-type structure demonstrate high hardness, and chemical and thermodynamic stability, with metal-like electrical and thermal properties. This features allowed for their utilization as advanced surface protective coatings for tribological[2] or biomedical applications[3], or diffusion barriers for metal interconnects in semiconductor devices[4] or an alternative to gold and silver in plasmonic applications [5].

Attempts to improve properties of AlN and TiN via different material engineering strategies (e.g. alloying, defect tuning, nanocomposite formation) often require a reliable determination of their electronic structure, surface composition, chemical state and local chemical environment. Although all latter can be revealed by employing X-ray photoelectron spectroscopy (XPS), the lack of reliable references as well as suitable charge references and surface band bending in AlN significantly complicates the analysis and interpretation of the results. Since the vast majority of laboratory XPS measurements are performed *ex situ*, additional complications arise due to surface contaminations and oxidation. Even further sputter-etching of the surface using Ar ions (Ar$^+$) in an XPS chamber can additionally modify the surface chemistry and lead to misleading results.

The modified Auger parameter (AP), α', has proven to become an effective tool for chemical state and local chemical environment analysis for a broad range of conductive and semiconducting materials [6–10]. AP of an element in a compound is defined as the sum of kinetic energy (KE) of a sharp core-core-core level type or core-core-valence level type Auger transition (KE (XYZ)) and binding energy (BE) of the core-level photoelectron involved in the Auger transition (BE (X/Y/Z))[11]:

$$\alpha' = KE\ (XYZ) + BE\ (X/Y/Z) \qquad (1)$$

Since AP is determined from the relative position of Auger and photoelectron peaks it is insensitive to static charging effects and inaccurate energy scale calibration. The Auger parameter is typically presented using Wagner plots. A Wagner plot is a scatter plot in which each point corresponds to the BE of a core-level electron and the KE of a corresponding Auger transition along with the calculated AP. AP shifts as a result of modification of the valence charge of the core-ionized atom ($\Delta R_a$) or due to differences in the extra-atomic relaxation energy ($\Delta R_{ea}$), described as following [12]:

$$\Delta \alpha' = 2 (\Delta R_a + \Delta R_{ea}) \quad (2)$$

Therefore, any impact from the changes in the oxidation state or electronegativity of the nearest-neighbors on $\Delta R_a$ can be efficiently determined by calculating AP. Moreover, any deviations in coordination number, the distance and electronic polarizability of the core-ionized atom's nearest-neighbors can affect $\Delta R_{ea}$ and therefore alter AP[13]. Extreme sensitivity to changes in local chemical environment due to double core-hole final state in the Auger process makes Auger parameter very sensitive to any surface or bulk chemical modifications. Due to latter, the exposure to ambient air or $Ar^+$ sputter-etching, commonly inducing surface modifications, can affect the AP of a given element. This especially holds for N $KL_{2,3}L_{2,3}$ and Ti $L_{2,3}M_{2,3}V$ Auger emission lines as the probing depth of these Auger electrons is less than 3 nm (see below).

The most studied and most intense are KLL and LMM Auger transitions for majority of elements, require excitation of a photoelectron from K and L core shells, respectively. Owing to the high BE of 1s electrons in Al (~1559 eV) and Ti (~4966 eV) containing compounds excitation of corresponding Auger transitions can be realized only using hard X-rays. In this regards, state-of-the-art lab-based hard X-ray photoelectron spectroscopy (HAXPES) with x-ray energies higher than 2 keV complements conventional soft X-ray based XPS[14,15]. Although the higher energy of X-ray source allows to excite deep core-level photoelectrons for triggering the associated deep core–core Auger transition, photo- and Auger electrons are emitted from different depth due to dependence of the former KE and independence of the latter KE from the X-ray excitation energy. This is illustrated in Fig.1 for TiN- and AlN-containing elements as the probing depth of the Al, Ti, N core-lines and Auger transitions for soft

Al-Kα (1 486.7 eV) and hard Cr-Kα (5 414.7 eV) X-rays used in the present study. The photoelectrons and corresponding Auger electrons involved in Auger parameter calculation originate from different depths. This becomes critical for the chemical state analysis of semiconducting materials as depth-dependent shifts in Fermi level may occur due to differential charging or band bending[16]. However, by combining soft Al-Kα (XPS) and hard Cr-Kα (HAXPES) under adequate calibration routines the probing depth of photoelectrons and corresponding Auger lines can be matched[15]. In this regard, AP study for AlN thin film at constant probing depth can be realized by probing Al 2p and Al 2s by means of XPS and Al KLL Auger lines by means of HAXPES. For TiN a set of APs can be determined by measuring Ti $2p_{3/2}$ and Ti $L_{2,3}M_{2,3}V$ by XPS and Ti $2p_{3/2}$ and Ti $KL_{2,3}L_{2,3}$ and $KL_{2,3}M_{4,5}$ by HAXPES. N 1s and N $KL_{2,3}L_{2,3}$ for both materials can be probed by XPS satisfying the minimum difference in the probing depth. It is important to note that Auger electrons and photoelectrons originating from the first 5-10 nm of the near-surface area are more susceptible to the impact from surface adsorbates, oxidation upon exposure to ambient air and Ar-ion-induced surface modifications. Therefore, surface modification can have different impact on a given pair of photo- and Auger electrons used for AP calculation depending on their depth of origin. The latter motivates selecting Auger and photoelectrons originating from the same depth of the sample.

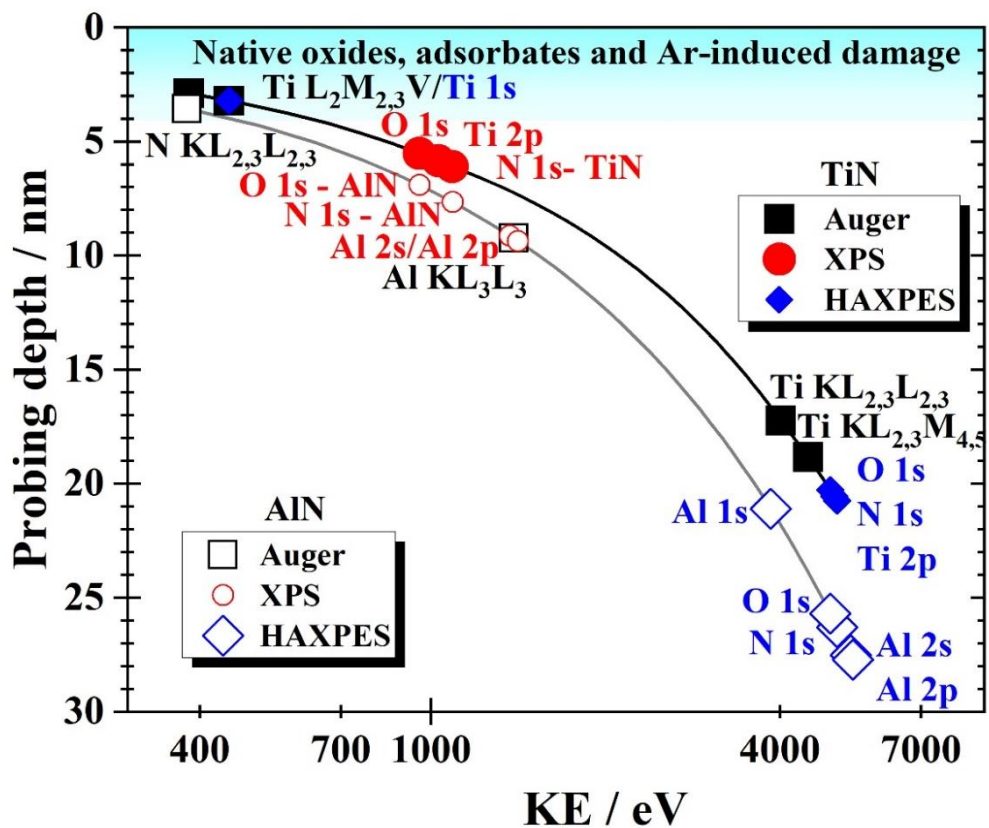

**Fig. 1.** Probing depth of photo- and Auger electrons excited by Al-Kα (XPS) and Cr-Kα (HAXPES) X-rays for elements constituting AlN and TiN calculated as a function of their kinetic energy at the emission angle of 0° deg.

Although reports on the AP of TiN and AlN are available in the literature they are usually acquired *ex situ* [8,17] or calculated considering photoelectrons and Auger electrons with different information depth[18]. APs determined in such a way are not reliable references for future studies. This is first of all related to possible differential charging effects in case of insulating AlN thin films[16]. Moreover, the charge referencing in the latter works was performed by using C 1s binding energy in adventitious carbon, which is widely known to be an inappropriate charge reference[19], affecting the accuracy of the reported values. Considering the increasing relevance of the AP for the analysis of electronic and chemical properties of different compounds, determination of reliable references for indisputably technologically important AlN and TiN thin films is of primary importance for further development and advancement of these materials and their applicability. Moreover, the lack of systematic study on the effects of air-exposure and Ar$^+$ sputter-cleaning-induced surface modification on AP limits understanding of the intrinsic properties of these industrially-relevant materials.

Here we perform a comprehensive Auger parameter analysis of magnetron sputter-deposited AlN and TiN thin films using a combination of XPS and HAXPES techniques. We characterize AlN and TiN thin films *in situ*, after exposure to ambient air and after subsequent sputter-etching by Ar$^+$. We show how different surface conditions can impact surface- and bulk-sensitive Auger parameters and subsequent interpretation of the results.

**Experimental**

TiN and AlN thin films are deposited by means of direct current (DC) reactive magnetron sputtering of Al and Ti targets (2" diameter and 99.9 % purity). The deposition of TiN thin films was carried out in reactive atmosphere of argon (18 sccm) and nitrogen (8 sccm) ensuring 0.5 Pa pressure at 250 °C substrate temperature. TiN films are grown by applying a moderate RF substrate bias with a frequency of 13.56 MHz and a potential of -75V to ensure low oxygen content. TiN thin films are grown on Si substrates. The deposition of AlN thin films was carried out in reactive atmosphere of argon (20 sccm) and nitrogen (10 sccm) ensuring 2.25 mTorr pressure at 300 °C substrate temperature. AlN thin films are grown on glass substrate for XPS/HAXPES study to avoid differential charging effects. $\Theta$-2$\Theta$ X-ray diffraction scans (not shown here) indicate that TiN and AlN are single phase fcc-NaCl structured and hexagonal-structured, respectively[20,21].

XPS and HAXPES measurements are performed using a Physical Electronics Quantes spectrometer equipped with monochromated Al-K$\alpha$ and Cr-K$\alpha$ X-ray sources. For each materials system, three sets of core-level and Auger spectra are acquired: one from as-grown *UHV*-transferred samples, from air-exposed samples and Ar$^+$ sputter-etched samples. UHV-transfer is performed by transferring the samples from deposition chamber to XPS/HAXPES analysis chamber through a Physical Electronics UHV transfer system allowing sample transfer at pressures below 5×10$^{-7}$ Pa[22]. Since the sample is not exposed to ambient air after deposition, we call it *in situ* sample. The samples are exposed to ambient air after *in situ* measurements to study the effects of surface contaminations and room temperature oxidation on the calculated AP. The effect of Ar$^+$ sputter-etching is studied by performing sputter-cleaning of the air-exposed samples using 1 keV Ar ion beam exposed at 2×2 mm$^2$ area of the samples at 45° incidence angle. The measurements of Auger and photoelectron lines are performed at pass energy of 69 eV with 0.125 eV measurement step and electron emission angle of 0° with respect to the surface normal to facilitate a high count rate. The pressure during the acquisition is below 5 ×10$^{-6}$ Pa. Charge neutralization for the measurements on AlN is achieved using a dual beam charge neutralization system based on a low-energy electron flood gun and a low energy positive ion source. The energy scale of the instrument is calibrated using an effective two-point calibration using Au 4f$_{7/2}$ photoemission lines acquired from the reference sample of gold using both X-ray sources that allows to create reference points at both ends of the KE scale. All peak positions are therefore shifted by ΔE value determined as a difference between Au 4f$_{7/2}$ peak position obtained from sputter-cleaned gold reference sample and ISO recommended standard BE of Au 4f$_{7/2}$, 83.96 eV[23]. The difference in static charging for XPS and HAXPES measurements is compensated by aligning the N1 1s and Al 2p for TiN and AlN thin films, respectively, acquired by Cr-Kα so that they match the values measured by Al-Kα, and shifting all Cr-Kα spectra accordingly. Spectra analysis is performed in Casa XPS software. Shirley background subtraction is used for all spectra. Quantification analysis is carried out using relative sensitivity factors supplied by the instrument manufacturer. The Auger and photoelectron peak positions are determined by the average of two cubic functions fitted by shifting the fitting region by 0.4-0.5 eV. Due to low signal to noise ratios all Auger emission lines are smoothed using Savitzky-Golay filtering method fitting over 15-25 adjacent data points with a 3$^{rd}$ order polynomial.

*TiN charge reference*. All spectra are referenced to the Fermi level (FL) edge cut off, which defines the 0 eV on the binding energy scale[23]. The Fermi level edge, clearly observed for all TiN layers as the drop in the density of states (Fermi edge), is set to 0 eV after fitting by the ''Step Down'' background type in Casa XPS and all spectra are shifted accordingly if needed.

***AlN charge reference***. Spectra from AlN thin films do not exhibit FL cut-off due to their semiconducting character and there is no proper and consistent binding energy scale reference reported in the literature. In our work, the reference Al 2p BE for AlN is determined by applying the following procedure. AlN thin films are grown on Si and borosilicate glass substrates under conditions described above. Immediately after AlN growth, Au-decoration of as-grown AlN thin films is performed by means of magnetron sputtering of Au target in Ar atmosphere for 5 seconds to ensure gold particle decoration rather than the formation of a closed Au film. An electronically floating surface is verified as the absence of photoelectron signal from Au-decorated AlN thin film grown on insulating glass substrates connected to the spectrometer stage by metallic clips. After Au decoration the sample is transferred *in situ* using a UHV transfer chamber from the deposition chamber to spectrometer. The samples are stored at a pressure below $10^{-5}$ Pa throughout the transport process. XPS measurements are performed using a Physical Electronics Quantera spectrometer equipped with monochromated Al-K$\alpha$ source. The base pressure during the spectra acquisition is below $10^{-6}$ Pa. Charge neutralization is achieved using a dual beam charge neutralization system based on a low-energy electron flood gun and a low energy positive ion source. For charge correction, all peak positions are thereafter shifted ensuring that the middle portion of the drop in the density of states at the FL (Fermi edge) coincides with 0 eV on the binding energy scale. Al 2p core-level photoelectron binding energy determined in this way serves as a charge reference for samples analyzed by HAXPES/XPS.

Inelastic mean free paths ($\lambda$) of photo-emitted and Auger electrons in AlN and TiN thin films for determination of the probing depths ($3\times\lambda\cos(\alpha)$, where $\alpha$ is the photo- and Auger electron detection angle with respect to the surface normal) are calculated using Quases software package, based on the Tanuma, Powell, Penn formula (TPP-2 model)[24].

## Results

***Auger parameter of TiN***

Fig. 2 shows core-level photoelectron and Auger emission spectra for TiN thin films acquired *in situ* in their as-deposited state as well as after exposure to ambient air and after subsequent Ar$^+$ sputter-etching using a combination of XPS/HAXPES. The corresponding peak positions are given in Table 1. The survey spectra are shown in Supplementary Fig. 1A. A general feature observed in all spectra is a significant decrease of signal intensity upon air-exposure due to attenuation of the signals by surface contaminations and oxides. However, the intensity is restored after Ar$^+$ sputter-etching and even exceeds the intensity of the spectra acquired *in situ* in some cases. Ti 2p spectrum from pristine TiN thin films measured *in situ* using Al Kα, Fig. 2A, demonstrates features typical for TiN with satellite peaks on the high BE side of the main line [25–27]. The corresponding N 1s spectrum, Fig. 2B, contains a low intensity peak on the low BE side of the main line implying the formation of titanium oxynitrides on the surface even during *in situ* measurements. Upon air exposure, the intensity of oxynitride and oxide peaks in N 1s and Ti 2p spectra increases while the intensity of the main line decreases that is in agreement with previous studies[28]. After Ar$^+$ sputter-etching, the oxide and oxynitride components of the Ti 2p spectrum are strongly reduced and the corresponding oxynitride peak on the low BE side of the main line of the N 1s spectrum nearly disappears. However, the Ti 2p$_{3/2}$ peak maximum shifts towards lower BE side due to the formation of N-depleted TiN bonds resulting from preferential sputtering. N 1s and Ti 2p spectra measured with Cr Kα remain unchanged regardless of surface condition. Moreover, the peaks contributions from the oxides and oxynitrides are not observed in the N 1s and Ti 2p spectra measured with Cr Kα compared to Al Kα. This suggests that surface oxides and contaminations layers do not significantly contribute to the Cr Kα signal, due to the greater information depth for Ti 2p and N 1s photoelectrons with Cr Kα (see Fig. 1). Importantly, the Ti 2p spectra acquired by Cr Kα are shifted to lower binding energy compared to those measured by Al Kα. This shift may be attributed to a slightly higher O content in the near-surface regions and less oxygen in the bulk of the film. This is reflected in a higher Ti 2p$_{3/2}$-N 1s difference, an indicator of a charge transfer from Ti to N[26,29], for Al Kα than for Cr Kα. This difference increases for both XPS and HAXPES upon air exposure of the TiN films (Table 1).

N $KL_{2,3}L_{2,3}$ and Ti $L_{2,3}M_{2,3}V$ Auger emission lines excited by both Al Kα and Cr Kα (Fig. 2C, D) show only minor shifts (≤0.2 eV), even though they originate from the near-surface region, which is subjected to oxidation and $Ar^+$ bombardment (Fig. 1). The most pronounced peak shifts are observed after $Ar^+$ sputter-etching when compared to *in situ* results. Importantly, N $KL_{2,3}L_{2,3}$ overlaps with Ti $L_3M_{2,3}M_{2,3}$ but the nitrogen Auger line is more intense and therefore remains suitable for peak maximum determination. Ti $L_3M_{2,3}M_{2,3}$ Auger line is therefore omitted in the present study to avoid ambiguity associated with Auger line deconvolution and fitting. Instead, the strong Ti $L_{2,3}M_{2,3}V$ Auger line is used for AP calculation. Ti $KL_{2,3}M_{2,3}$ and Ti $KL_{2,3}L_{2,3}$ Auger lines are less surface sensitive than N $KL_{2,3}L_{2,3}$ and Ti $L_{2,3}M_{2,3}V$ (Fig. 1) but also exhibit a small peak shift upon air-exposure and $Ar^+$ sputter-etching (Fig. 2E, F, Table. 1). The shift of Ti $KL_{2,3}L_{2,3}$ Auger lines (Table 1, Fig. 2E) first towards low KE side, due to oxidation upon air exposure, and then to high KE side due to preferential sputtering of N, is in agreement with previous studies[30].

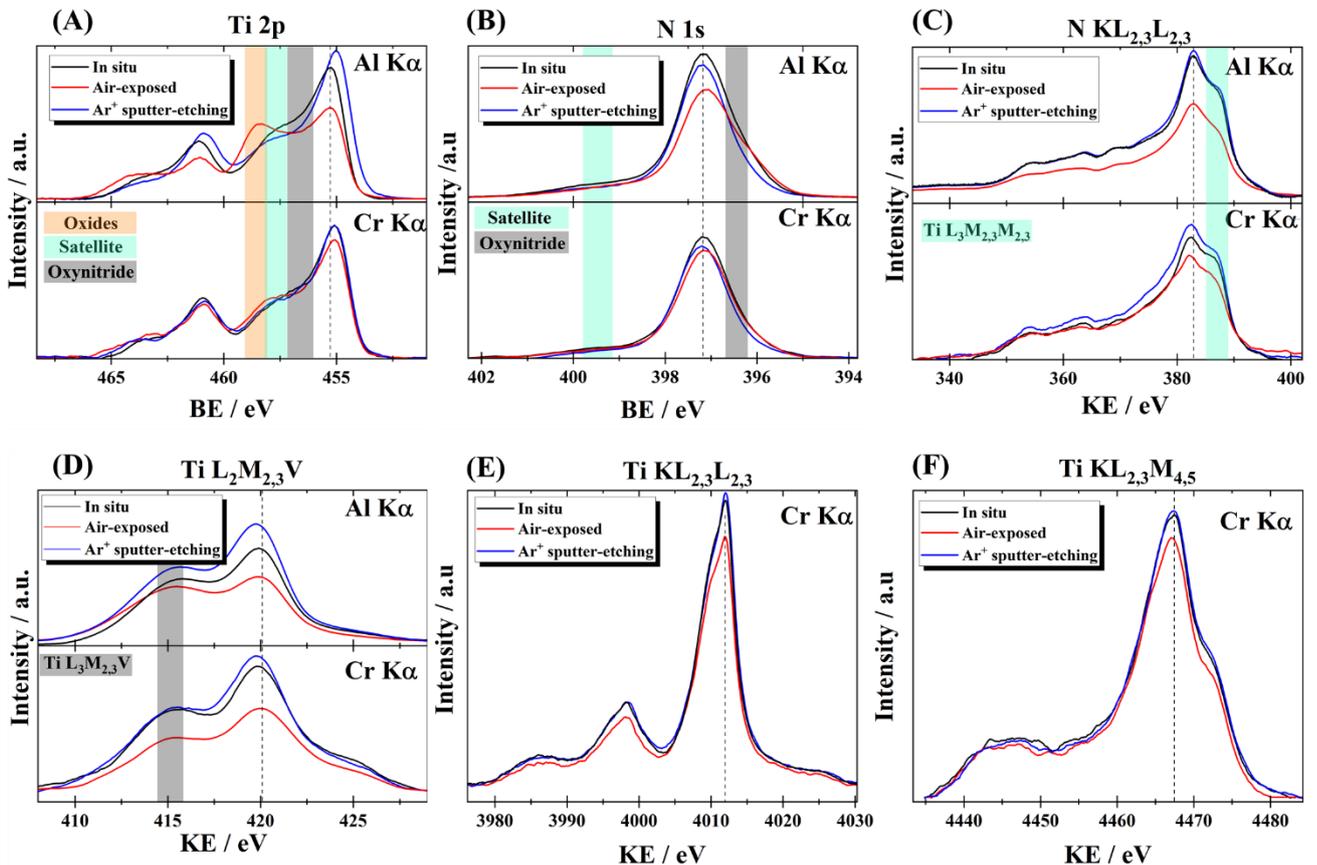

**Fig. 2.** Summary of XPS and HAXPES spectra of Ti 2p (A) and N 1s (B) core-levels and N KL$_{2,3}$L$_{2,3}$ (C), Ti L$_{2,3}$M$_{2,3}$V (D), Ti KL$_{2,3}$L$_{2,3}$ (E), and Ti KL$_{2,3}$M$_{4,5}$ (F) Auger emission lines for TiN thin films grown on Si (001) substrates measured *in situ*, after air-exposure and after subsequent Ar$^+$ sputter-etching with 1 keV beam. The conductive samples are grounded during XPS and HAXPES measurements. Black dashed lines denote the position of the peaks for *in situ* measured samples using Al Kα, while in case of Ti KL$_{2,3}$L$_{2,3}$ and Ti KL$_{2,3}$M$_{4,5}$ Auger lines the black dashed line denotes the position of the peaks for *in situ* measured samples using Cr Kα. All peak positions acquired by Cr Kα are shifted to match the BE of N 1s to the values measured by Al Kα.

**Table 1.** Core-level photoelectron BEs and Auger electron KEs acquired using XPS (Al Kα) and HAXPES (Cr Kα) for TiN thin films measured *in situ*, after exposure to ambient air and after subsequent sputter-etching with 1keV Ar$^+$ together with calculated α'. Ti 1s core-level BEs are not used for calculation of the Auger parameter.

| | Al Kα | | | | | Cr Kα▼ | | | | | Auger parameter | | | |
|---|---|---|---|---|---|---|---|---|---|---|---|---|---|---|
| | Ti 2p$_{3/2}$ / eV | N 1s / eV | N KL$_{2,3}$L$_{2,3}$ / eV | Ti L$_{2,3}$M$_{2,3}$V / eV | Ti 2p-N1s / eV | Ti 1s / eV | Ti KL$_{2,3}$L$_{2,3}$ / eV | Ti KL$_{2,3}$M$_{4,5}$ / eV | Ti 2p$_{3/2}$ / eV | Ti 2p-N1s / eV | α'$_N$ (N 1s, N KLL) | α'$_{Ti}$ (Ti 2p, Ti LMV) | α'$_{Ti}$ (Ti 2p, Ti KLL) | α'$_{Ti}$ (Ti 2p, Ti KLM) |
| **In-situ** | 455.29 | 397.17 | 382.91 | 420.10 | 58.10 | 4966.04 | 4011.75 | 4467.24 | 455.07 | 57.90 | 780.08 | 875.37 | 4466.82 | 4922.32 |
| **Air-exposed** | 455.32 | 397.13 | 382.86 | 420.06 | 58.13 | 4966.12 | 4011.70 | 4467.13 | 455.09 | 57.96 | 779.99 | 875.33 | 4466.79 | 4922.22 |
| **Ar$^+$ sputter-etching** | 455.07 | 397.19 | 382.88 | 419.91 | 58.08 | 4965.80 | 4012.05 | 4467.34 | 455.07 | 57.88 | 780.07 | 875.18 | 4467.12 | 4922.40 |

▼All peak positions measured by Cr Kα are shifted so matching N1s BE to the value measured by Al Kα.

Table 2. XPS-derived quantitative analysis for TiN thin films measured *in situ*, after air exposure, and after Ar ion sputter-etching. Typical uncertainties in the measurement of elemental concentrations are ±2 at.%

|  | Ti / at.% | N / at.% | C / at.% | O / at.% |
|---|---|---|---|---|
| **In situ** | 39.3 | 54.3 | 1.0 | 5.4 |
| **Air-exposed** | 31.3 | 39.1 | 7.3 | 22.3 |
| **Ar$^+$ sputter-etching** | 38.9 | 43.6 | 1.2 | 16.3 |

Wagner plots for TiN containing elements are shown in Fig. 3 to summarize the relationship between core-level photoelectron BEs and Auger electron KEs and to illustrate the impact of air exposure and Ar$^+$ sputter-etching on AP. Importantly, photoelectron lines and Auger emission lines for Auger parameter calculation were selected considering the following criteria:

(i) core-level photoelectrons should participate in the Auger transition to ensure the AP calculation reflects the correct physical processes and

(ii) Auger electrons and photoelectrons should exhibit similar information depth to minimize differential charging effects.

Based on these criteria, we exclude $\alpha'_{(Ti\ 1s,\ Ti\ KL_{2,3}L_{2,3})}$ from the analysis because Ti KL$_{2,3}$L$_{2,3}$ Auger electrons and Ti 1s photoelectrons originate from significantly different depths (Fig. 1). Following (ii), we exclude $\alpha'_{(Ti\ 1s,\ Ti\ L_{2,3}M_{2,3}V)}$ because although Ti 1s and Ti L$_{2,3}$M$_{2,3}$V originate from the same depth but Ti 1s electrons belong to K shell of the Ti atom.

Intriguingly, $\alpha'_{N(N\ 1s,\ N\ KL_{2,3}L_{2,3})}$ remains unchanged within the measurement uncertainty upon air exposure or Ar$^+$ sputter-etching. While exposure to air does not significantly affect $\alpha'_{Ti(Ti\ 2p_{3/2},\ Ti\ L_{2,3}M_{2,3}V)}$, Ar$^+$ sputter-etching results in $\alpha'_{Ti(Ti\ 2p_{3/2},\ Ti\ L_{2,3}M_{2,3}V)}$ shift towards values of pure titanium (872.8 ± 0.4 eV [31]) as the nearest-neighbor environment around Ti atoms is modified by preferential sputtering of N making the surface more metal-like. The bulk-sensitive $\alpha'_{Ti(Ti\ 2p_{3/2},\ Ti\ KL_{2,3}L_{2,3})}$ measured using Cr Kα remains unaffected upon air exposure and Ar$^+$ sputter-etching, Fig. 3C. However, oxidation of TiN surface via exposure to air results in the decreased of $\alpha'_{Ti(Ti\ 2p_{3/2},\ Ti\ KL_{2,3}M_{4,5})}$. Whereas Ar$^+$ sputter-cleaning of the TiN surface causes the increase of $\alpha'_{Ti(Ti\ 2p3/2,\ Ti\ KL2,3M4,5)}$. This is surprising since both Ti $2p_{3/2}$, Ti $KL_{2,3}M_{4,5}$ are bulk sensitive as $KL_{2,3}L_{2,3}$. However, this shift of ±0.1 eV is rather marginal for Auger parameter and lies within the measurement uncertainty.

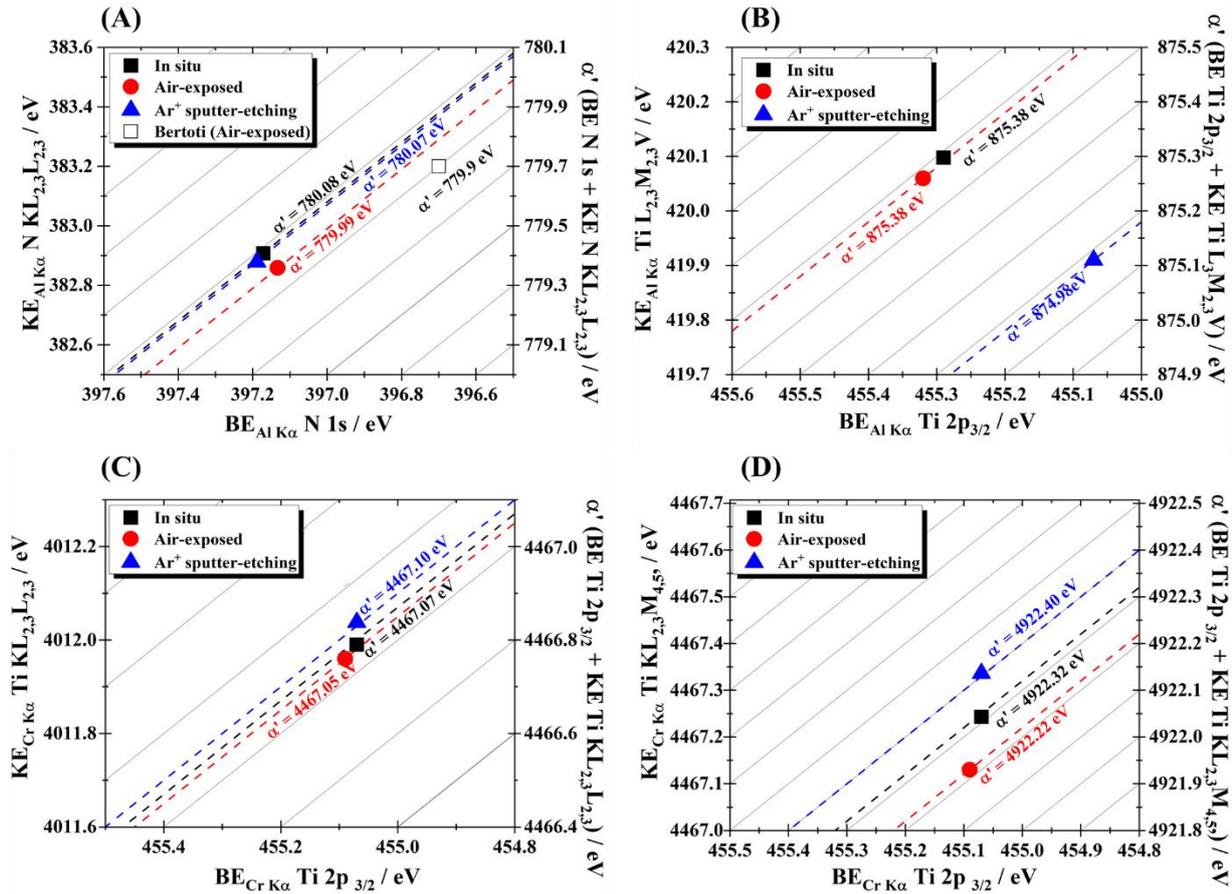

**Fig. 3.** Wagner plots depicting $\alpha'_{N(N\ 1s,\ N\ KL2,3L2,3)}$ (A), $\alpha'_{Ti\ (Ti\ 2p3/2,\ Ti\ L2,3M2,3V)}$ (B), $\alpha'_{Ti\ (Ti\ 2p3/2,\ Ti\ KL2,3L2,3)}$ (C), $\alpha'_{Ti\ (Ti\ 2p3/2,\ Ti\ KL2,3M4,5)}$ (D) for TiN thin films measured *in situ*, after air exposure and after subsequent Ar$^+$ sputter-etching. $\alpha'_{N\ (N\ 1s,\ N\ KL2,3L2,3)}$ reported by I. Bertoti[8] for air-exposed TiN thin films charge-referenced to C 1s of adventitious carbon is provided as a reference.

*AlN thin films*
*Charge reference*

An accurate XPS analysis of AlN thin films is hindered by static charging due to insulating nature of the material. This necessitates the use of a reliable charge reference.

Since referencing to BE of C1s in adventitious carbon and Ar 2p of implanted argon are not the accurate methods for charge referencing [32,33], we employ the noble metal decoration method[34]. We decorate AlN thin films by sputtering Au on AlN thin films while avoiding the formation of a closed Au thin film. Fig. 4 exhibits core-level spectra, Auger emission lines and valence band spectra recorded from AlN thin films decorated with Au and UVH-transferred to XPS spectrometer. Peak positions are determined by fitting all core-level applying constraints for the spin orbit peaks of Au 4f as 3:4 for Au $4f_{5/2}$ and Au $4f_{7/2}$, respectively, and for the spin orbit peaks of Al 2p as 1:2 for Al $2p_{1/2}$ and Al $2p_{3/2}$, respectively. The position of Fermi edge is determined by calculating the derivative of the signal recorded in the vicinity of the Fermi level, Fig. 4A (bottom). For charge correction, the position of Fermi edge is set to 0 eV, and all other peaks are shifted respectively. After this charge correction, the BE of Au $4f_{7/2}$ photoelectrons, Fig. 4B, is measured at 83.89 eV, close to the ISO value of 83.96 eV[23]. This slight discrepancy can be assigned to the cluster size-dependent BE of Au $4f_{7/2}$ electrons when the gold surface coverage is below one monolayer[35–37]. Here we chose the Fermi edge as a reference as it is not size dependent. Owing to this charge correction, the Al 2p, Al 2s and N 1s photoelectron BEs as well as N AP of AlN thin film are determined (Fig. 4C-F and Table 3). Al 2p and N 1s core-level photoelectron BE determined in in this work are in a good agreement with the BE of 74.11 eV and 397.2 eV, respectively, determined previously for *in situ* Cr-capped AlN thin films[38]. The Al $2p_{3/2}$ BE of AlN determined in this way is used as a reference throughout the rest of this manuscript.

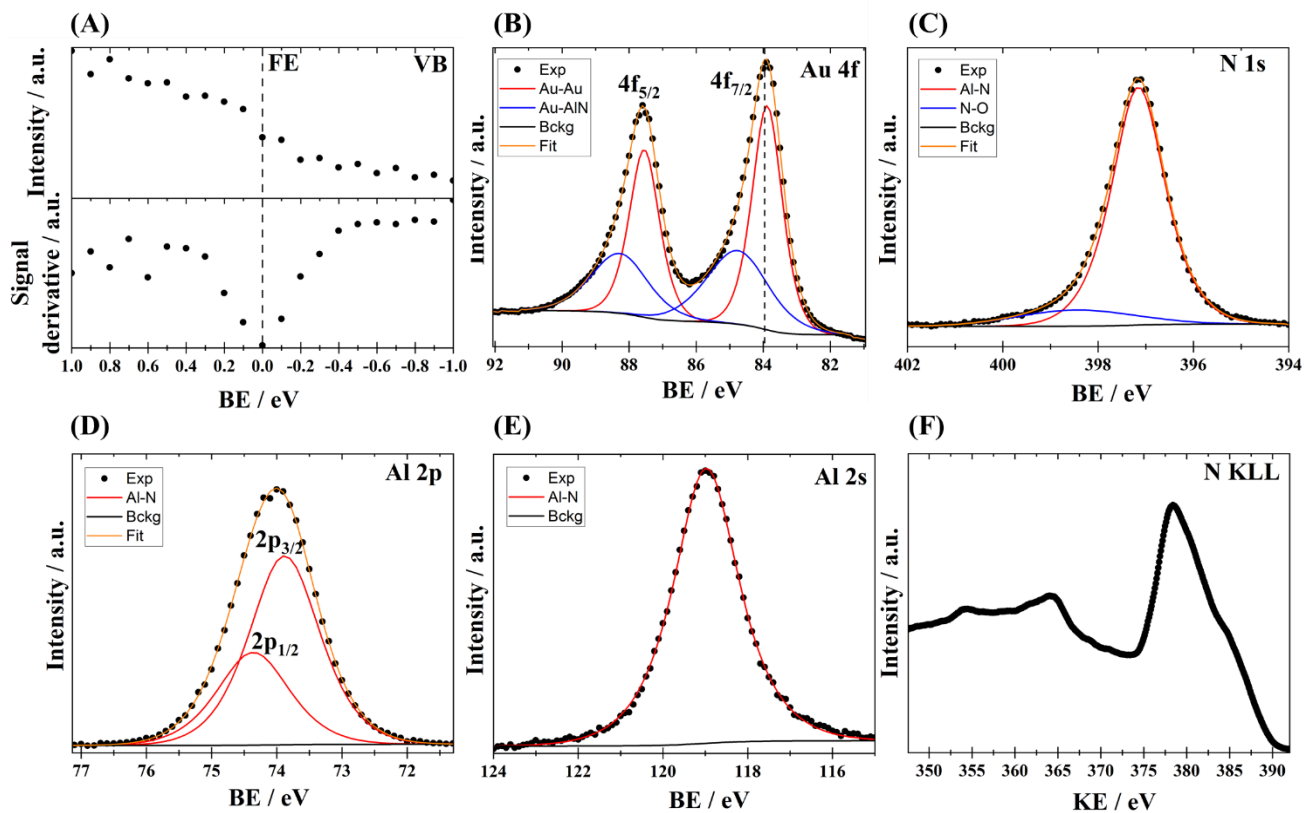

**Fig. 4.** Summary of XPS spectra for Au-decorated AlN thin films grown on Si (001) substrates and transferred to XPS spectrometer under UHV conditions: the spectra recorded in the vicinity of the Fermi level (A, top) together with the signal derivative (A, bottom), Au 4f (B), N1s (C), Al 2p (D), Al 2s (E) core-level spectra and N KL$_{2,3}$L$_{2,3}$ (C) Auger emission line. Black dashed line in A denotes the standard BE the Au 4f$_{7/2}$ electrons in bulk gold sample according to the ISO 15472:2010 document for monochromatic Al Kα sources.

*Auger parameter of AlN*

Fig. 5 shows core-level photoelectron and Auger emission spectra for AlN thin films acquired *in situ*, after exposure to ambient air and after subsequent Ar$^+$ sputter-etching using a combination of XPS/HAXPES. The survey spectra are shown in Supplementary Fig. 1B. The corresponding peak positions are given in Table 3. The films are grown on insulating glass substrates to minimize differential charging effects. The charge referencing is performed by aligning all Al 2p$_{3/2}$ peaks acquired by Al Kα to Al 2p$_{3/2}$ BE determined for Au-decorated AlN thin films, as described in previous section. This referencing procedure also results in the Al 2s position being set in good agreement with the Al 2s BE determined for the Au-decorated sample (Table 3). In general, the spectra recorded using Cr Kα and Al Kα show a good agreement. No significant changes in peak shape or peak broadening can be observed for Al 2p and Al 2s spectra measured using both Cr Kα and Al Kα (Fig. 5A, B) after exposure to air or Ar$^+$ sputter-etching. The Al 1s core-level spectra shape and peak maxima position remain intact after exposure to air and after Ar$^+$ etching (Fig. 5C). The Al 2p spin-orbit splitting $2p_{1/2}$ - $2p_{3/2}$ component spacing ($\Delta_{1/2-3/2}$) is 0.48 eV for all samples, being in good agreement with the Au decorated reference sample. The N 1s peak shape does not change significantly upon air exposure, Fig. 5D, but the peak position shifts by ~ 0.1 eV to the lower BE side after exposure to air and by another ~ 0.1 eV after Ar$^+$ sputter-etching (Fig. 5D) relative to the *in-situ* sample. This shift is accompanied by the decrease of the N 1s-Al 2p difference implying the increase of the charge transfer from Al to N (Table 3).

Al $KL_{2,3}L_{2,3}$ spectrum of the *in situ* measured AlN thin film demonstrates a sharp and distinct peak (Fig. 4E) at 1388.17 eV that is in a close agreement with KE of Al $KL_{2,3}L_{2,3}$ transition of 1388.5 eV previously measured by the Bremsstrahlung component of the Mg radiation[39]. Upon air-exposure, see Fig. 4E, a distinct aluminum oxide component appears at the low KE side of the Al $KL_{2,3}L_{2,3}$ spectrum [39] while the peak position remain unchanged. In contrast, a shift towards higher KE side is observed after Ar$^+$ sputter-etching. Both air-exposure and Ar$^+$ sputter-etching also cause a shift of the N $KL_{2,3}L_{2,3}$ lines to higher KE side and the strongest impact has Ar$^+$ sputter-etching (Table 3).

Table 3. Core-level photoelectron BEs and Auger electron KEs acquired by XPS (Al Kα) and HAXPES (Cr Kα) for AlN thin films measured *in situ*, after exposure to ambient air and after subsequent Ar$^+$ sputter-etching together with calculated α'.

| | Al Kα | | | | | Cr Kα | | Auger parameter | | |
|---|---|---|---|---|---|---|---|---|---|---|
| | Al $2p_{3/2}$ / Al $2p_{1/2}$ / eV | Al 2s / eV | N $KL_{2,3}L_{2,3}$ / eV | N 1s / eV | N 1s-Al $2p_{3/2}$ / eV | Al $KL_3L_3$ / eV | Al 1s / eV | α'$_N$ (N 1s, N KL2,3L2,3) / eV | α'$_{Al}$ (Al 2s, Al KL2,3L2,3) / eV | α'$_{Al}$ (Al 2p3/2, Al KL2,3L2,3) / eV |
| **In-situ** | 73.88/74.37■ | 118.99 | 378.54 | 397.21 | 323.33 | 1388.61▼ | 1560.70 | 775.75 | 1507.66 | 1462.49 |
| **Air-exposed** | 73.88/74.36■ | 118.96 | 378.72 | 397.12 | 323.24 | 1388.57▼ | 1560.69 | 775.83 | 1507.53 | 1462.45 |
| **Ar$^+$ sputter-etching** | 73.88/74.36■ | 119.92 | 378.99 | 396.98 | 323.10 | 1388.82▼ | 1560.066 | 775.97 | 1507.73 | 1462.70 |
| **Au-decorated** | 73.88/74.36 | 118.98 | 378.64 | 397.16 | 323.25 | – | – | 775.58 | – | – |

■Al $2p_{3/2}$ is set to 73.88 eV as determined for Au-decorated AlN thin films on Si substrate and all other spectra acquired using Al Kα are shifted accordingly.

▼All peak positions measured by Cr Kα are shifted so matching Al 2p BE to the value measured by Al Kα.

Table 4. XPS-derived quantitative analysis for AlN thin films measured *in situ*, after air exposure and after subsequent Ar$^+$ sputter-etching. Typical uncertainties in the measurement of elemental concentrations are ±2 at.%

|  | Al / at.% | N / at.% | C / at.% | O / at.% |
| --- | --- | --- | --- | --- |
| **In situ** | 45.7 | 46.3 | 1.8 | 6.2 |
| **Air-exposed** | 41.8 | 35.1 | 3.5 | 19.6 |
| **Ar$^+$ sputter-etching** | 46.1 | 41.8 | 0.5 | 11.6 |

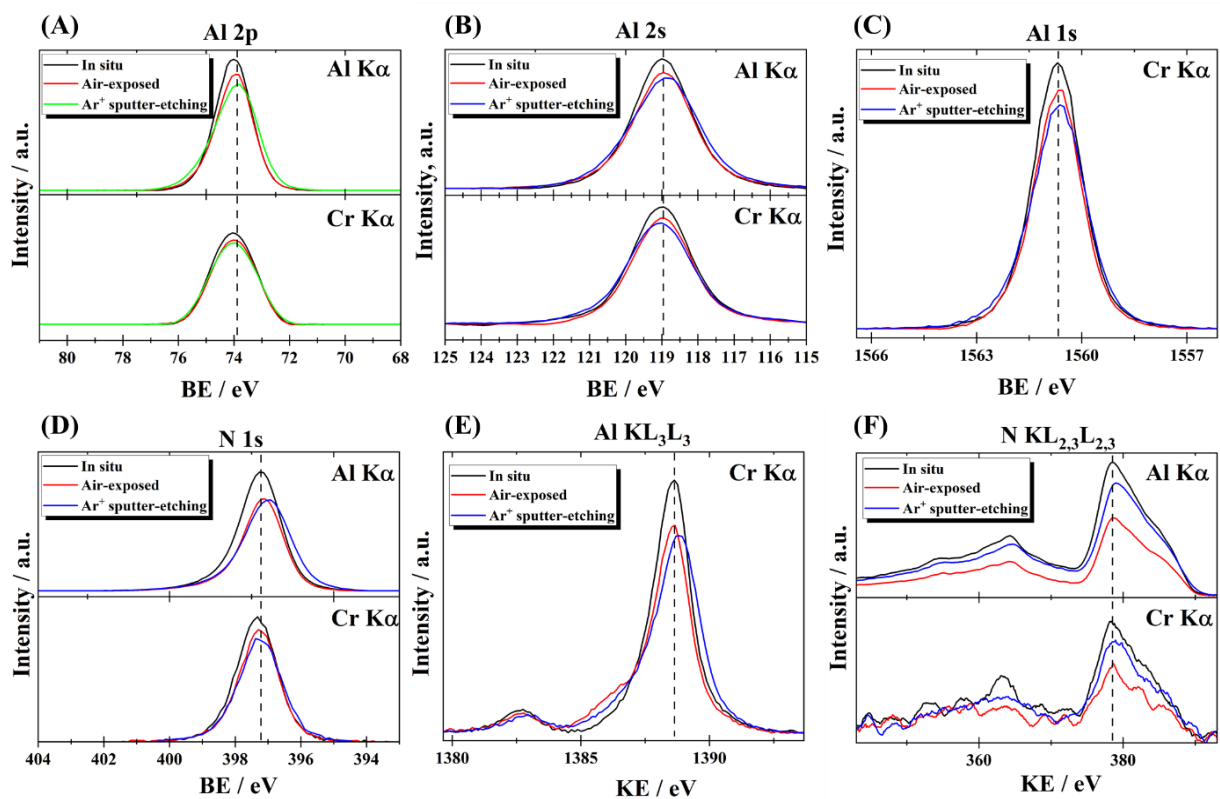

**Fig. 5.** Summary of XPS and HAXPES spectra of Al 2p (A), Al 2s (B) Al 1s (C), and N1s (D) core-levels and Al KL$_{2,3}$L$_{2,3}$ (E) and N KL$_{2,3}$L$_{2,3}$ (F) Auger emission lines for AlN thin films grown on insulating glass substrates measured *in situ*, after air-exposure and after subsequent Ar$^+$ sputter-etching with 1 keV beam. Black dashed lines denote the position of the peaks for *in situ* measured samples using Al Kα (and using Cr Kα for Al 1s and Al KL$_{2,3}$L$_{2,3}$). All peak positions acquired by Cr Kα are shifted to match the KE value of N KL$_{2,3}$L$_{2,3}$ Auger electrons to the value measured by Al Kα.

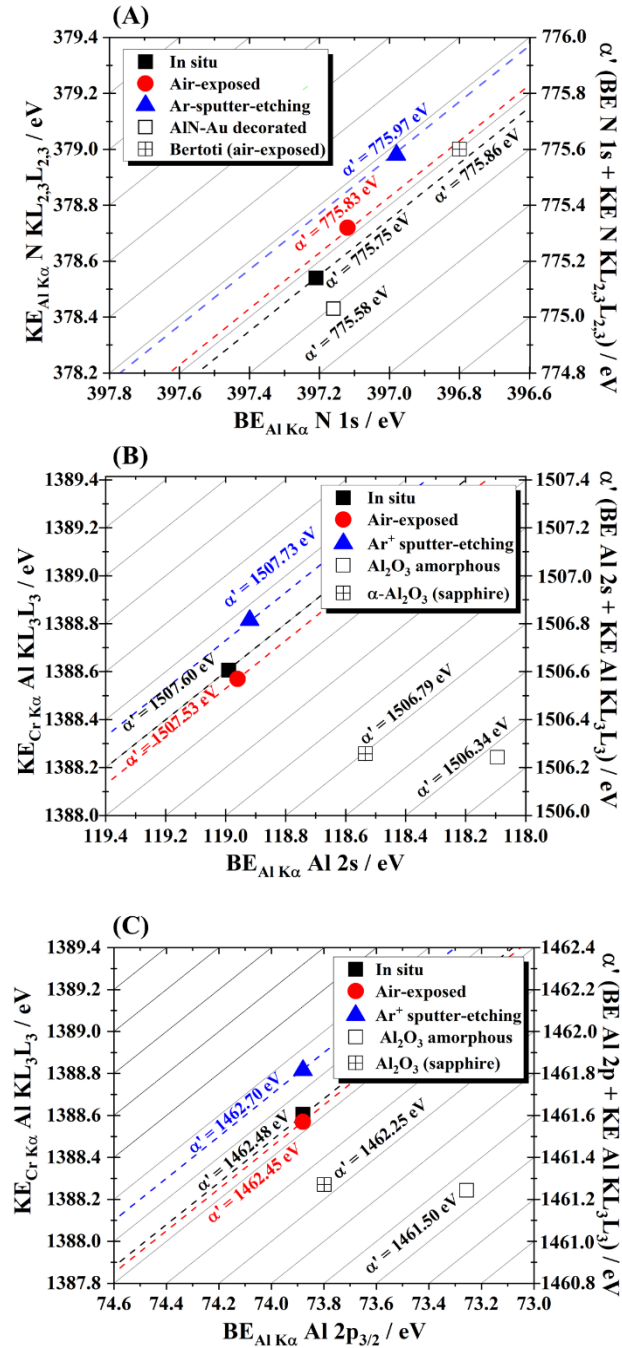

Fig. 5. Wagner plots depicting α'$_{N\ (N\ 1s,\ N\ KL2,3L2,3)}$ (A), α'$_{Al2s\ (Al\ 2s,\ Al\ KL2,3L2,3)}$ (B) and α'$_{Al2p\ (Al\ 2p,\ Al\ KL2,3L2,3)}$ for AlN thin films measured *in situ*, after air exposure and after subsequent Ar$^+$ sputter-etching. α'$_{N\ (N\ 1s,\ N\ KL2,3L2,3)}$ reported by I. Bertoti[8] for air-exposed AlN thin films charge-referenced to C 1s of adventitious carbon and for gold-decorated AlN thin films are provided as a reference in A. α'$_{Al\ (Al\ 2p,\ Al\ KL2,3L2,3)}$ and α'$_{Al\ (Al\ 2s,\ Al\ KL2,3L2,3)}$ for amorphous Al$_2$O$_3$ thin films and sapphire reported by S. Siol et al. [15] are provided as references in B and C.

20Pshyk *et al.* Empa 2024

The changes in core-level spectra and Auger emission spectra are reflected in the corresponding APs that are illustrated in Wagner plots, Fig. 5. Since Al $KL_3L_3$ Auger electrons and Al 1s photoelectrons originate from significantly different depths (Fig. 1) we exclude $\alpha'_{(Al\ 1s,\ Al\ KL3L3)}$ from the analysis. N AP of the pristine AlN thin films lies close to the previously reported value and agrees very well with the AP of Au-decorated AlN (Fig. 5A). The photoelectron BE and Auger peak positions reported by Bertoti at al.[8] are significantly different compared to the values determined in this work. Importantly, N AP shifts by +0.1 eV after air-exposure and further +0.1 eV after $Ar^+$ sputter-etching (Fig. 5A). Similar trend is observed for N1s-Al2p peak separation, a proxy for charge transfer from Al to N (Table. 3): the distance decreases by ~0.1 eV after exposure to air and then by ~0.2 eV after $Ar^+$ sputter-etching relative to the *in-situ* sample. Moreover, exposure to air does not change significantly both $\alpha'_{Al2s\ (Al\ 2s,\ Al\ KL2,3L2,3)}$ and $\alpha'_{Al2p\ (Al\ 2p,\ Al\ KL2,3L2,3)}$, Fig. 5B, C. However, the slight shift of +0.2 eV relative to the *in-situ* sample is observed for the latter APs after $Ar^+$ sputter-etching.

**Discussion**

The N AP of TiN does not significantly change in response to surface oxidation/contamination absorption or $Ar^+$ sputter-etching. In contrast, AlN exhibits a slight positive shift of approximately +0.2 eV. This is surprising given that both N 1s photoelectrons and N KLL Auger electrons originate from the near surface area of both compounds. However, the observed shift is small and lies within the experimental error for AP determination. Typical errors in the measurement of individual photoelectron peak positions are 0.l-0.2 eV, while for the broad Auger lines like N KLL the errors are about 0.2-0.4 eV. Therefore, N AP remains a reliable parameter for the characterization of both compounds after exposure to air and $Ar^+$ ion etching using XPS/HAXPES techniques. N AP for AlN and TiN thin film differs by 4.3 eV, which can be related to the different coordination number of N in these compounds (6-fold in NaCl structured face-centered lattice of TiN and 4-fold in wurtzite structured AlN) as well as to the specific electronic polarizability of the neighboring atoms around N anion in fcc and hex lattices.

Oxidation of TiN and AlN at room temperature proceeds via substitution of N by O leading to the formation of Ti-N-O or Al-N-O complexes with further formation of ultrathin (≤ 2 nm depending on exposure time and stoichiometry) surface oxide layers [40–43]. This is reflected in the decrease in the spectra intensity (Fig. 2, 5). It is well known that AlN forms a passivating surface oxide layer that prevents bulk oxidation, while TiN does not. Difference in oxidation of both compounds and its impact on the XPS and HAXPES spectra is reflected in the intensities of O 1s and C 1s core-levels spectra, Fig. 6. While the intensity of O 1s photoelectron lines measured by XPS for AlN thin films is ×3.5 higher after air exposure in comparison to the *in-situ* sample, the O 1s intensity from air exposed sample is only ×2.5 higher for the spectra measured by HAXPES. In contrast, for TiN thin films, the intensity of O 1s photoelectron lines measured by XPS is ×3.5 higher after air exposure in comparison to the *in-situ* sample, but this intensity is ×8.5 higher for the spectra measured by HAXPES. These results suggest that HAXPES can be effectively used to study air-transferred AlN films but is less suitable for TiN, where bulk oxidation is more pronounced. C 1s core-level spectra intensity significantly increases after air exposure when measured by XPS due to the formation of adventitious carbon on the surface of the films[32], which, in addition to oxide formation, contributes to the signal attenuation (Fig 6C, D). Importantly, the C 1s signal is not detected by HAXPES except for AlN thin films exposed to air, additionally highlighting surface insensitivity of HAXPES. Moreover, titanium oxynitride and oxide components appear in the Ti 2p spectra acquired by Al Kα and the low BE component in the N 1s spectra from the air exposed sample (Fig. 2B). However, these oxidation-related changes have a very minor impact on AP of Ti and Al, cf. Fig. 3 and Fig. 5. Ti $2p_{3/2}$ and Ti $L_{2,3}M_{2,3}V$ do not shift upon air-exposure and has no impact on the surface-sensitive AP ($\alpha'_{Ti\ (Ti\ 2p3/2,\ Ti\ L2,3M2,3V)}$). Similarly, the bulk-sensitive APs ($\alpha'_{Ti\ (Ti\ 2p3/2,\ Ti\ KL2,3M4,5)}$ and $\alpha'_{Ti\ (Ti\ 2p3/2,\ Ti\ KL2,3L2,3)}$) are mostly unaffected by surface oxidation (Table 1) due to the higher information depth of Ti $2p_{3/2}$ photoelectrons and Auger electrons acquired by Cr Kα as oxidation primary occurs in the near surface region. Similarly, Al 2p and Al 2s core-levels measured by Al Kα and the Al $KL_{2,3}L_{2,3}$ Auger line measured by Cr Kα are not affected by oxidation because the probing volume is larger in comparison to the surface oxide layer volume. Therefore, α'$_{Al2p\ (Al\ 2p,\ Al\ KL2,3L2,3)}$ and α'$_{Al2s\ (Al\ 2s,\ Al\ KL2,3L2,3)}$ also remain unchanged after the exposure of AlN thin film to ambient air. Assuming an oxide layer thickness ($d_{ox}$) of 2 nm for both materials, the fraction of the total signal originating from the oxidized surface—given by the expression $1 - \exp(-d_{ox}/\lambda)$, is approximately 40% for TiN and 20% for AlN, based on inelastic mean free path $\lambda$. As a result, Al 2p and $KL_{2,3}L_{2,3}$ peaks do not show peak shifts or broadening upon air exposure, while the Ti 2p spectrum does reveal a strong oxide component. This is supported by the fact that Al 2s and Al 2p peaks do not exhibit any peak shift or broadening upon exposure of the AlN sample to air while Ti 2p spectrum has a strong oxide component after exposure of TiN to air. This is also confirmed by a marginal difference in the Al 2p, Al 2s and N 1s peak positions measured by Al Kα and Cr Kα. They originate from different depths and the most surface sensitive lines, measured by Al Kα, remain intact after exposure to air. This indicates that the changes in the bonding environment of Al, Ti and N in the near-surface region of the material are not reflected in the AP. In AlN, the oxide/contamination layer is very thin contributing minimally to the recorded photoelectron and Auger lines. The formation of thin oxide layer is very common for many other nitride films after ambient transfer to the XPS/HAXPES[7,25,44,45]. For TiN, however, the oxidation is so strong giving rise to a well-defined photoelectron peak at the high BE side of the spectrum[46]. While this allows accurate determination of the photoelectron peak position, the broad nature of the Ti $L_{2,3}M_{2,3}V$ Auger line introduces uncertainty in peak maximum determination, making it challenging to assess oxidation effects on this Auger transition.

23Pshyk *et al.* Empa 2024

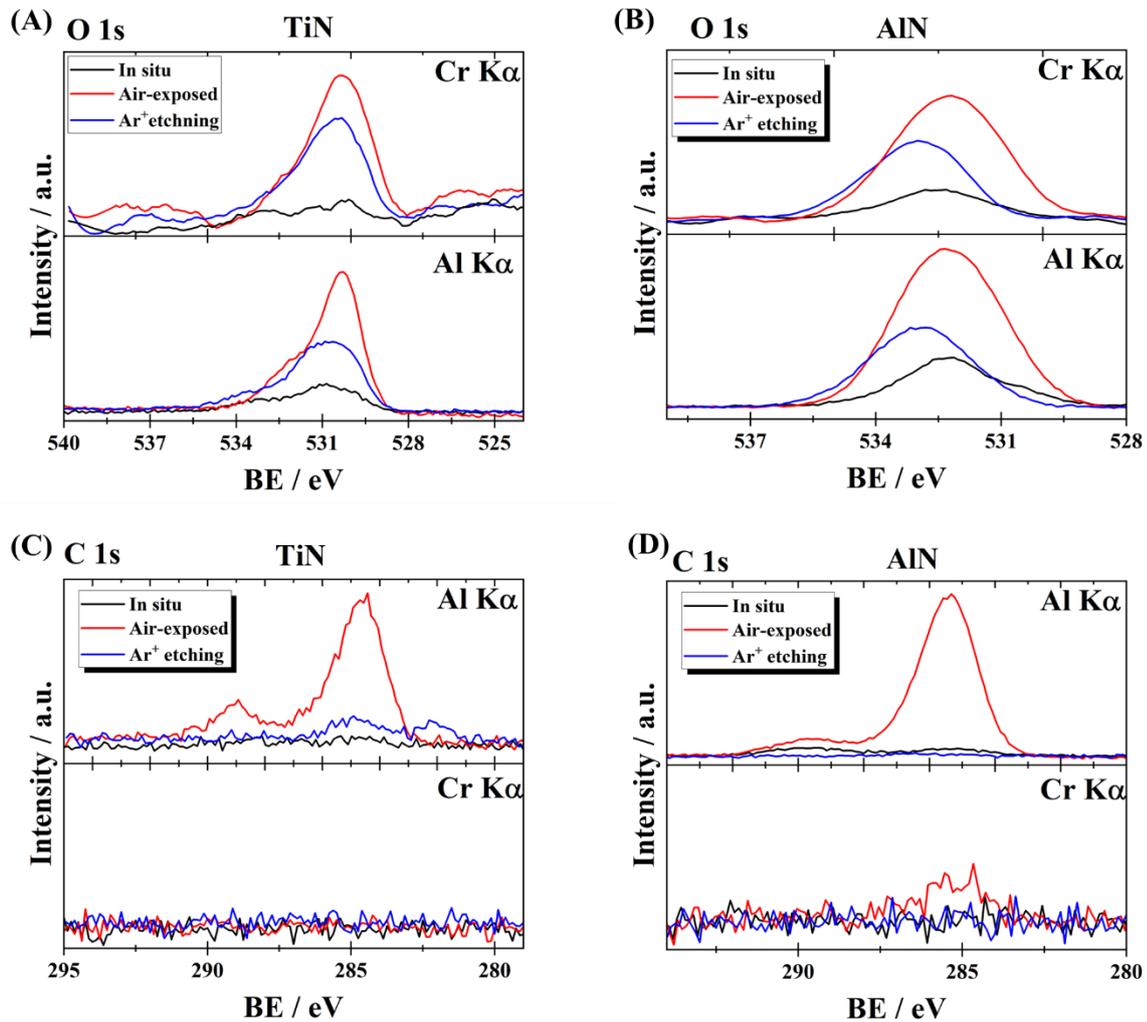

**Fig. 6.** O 1s and C 1s core-levels for TiN (A, C) and AlN (B, D) thin films grown on insulating glass substrates measured by XPS and HAXPES *in situ*, after air-exposure and after subsequent Ar⁺ sputter-etching with 1 keV beam.

Ar⁺ sputter-etching has the strongest impact on the AP of Al and α'$_{Ti\ (Ti\ 2p3/2,\ Ti\ L2,3M2,3V)}$ (Fig. 3, Fig. 5) and N in AlN. Although core-core-valence Auger transition related α'$_{Ti\ (Ti\ 2p3/2,\ Ti\ KL2,3M4,5)}$ and core-core-core Auger transition related α'$_{Ti\ (Ti\ 2p3/2,\ Ti\ KL2,3L2,3)}$ do not change within the measurement uncertainty, core-valence-valence Auger transition related α'$_{Ti\ (Ti\ 2p3/2,\ Ti\ L3M2,3V)}$ shifts toward the values of metallic Ti, 872.8 eV[31]. Furthermore, both core-core-core Auger transition related α'$_{Al\ (Al\ 2p,\ Al\ KL2,3L2,3)}$ and α'$_{Al\ (Al\ 2s,\ Al\ KL2,3L2,3)}$ shift towards AP of pure Al, 1465.9 eV[31]. This trend can be assigned to the formation of metal-rich surfaces in both cases due to Ar⁺ sputter etching. The preferential sputtering of lighter N occurs in this case because the mass difference between film constituents is large, e.g. for AlN $m_{Al}$=1.9 $m_N$, for TiN $m_{Ti}$=3.4 $m_N$. This leads to a large difference in the sputtering yield of the metal elements and nitrogen, here estimated from Monte-Carlo based TRIM [47] simulations: for Ti and N atoms it is 1.3 atoms/ion and 2.6 atoms/ion, respectively, while for Al and N atoms it is 1.5 atoms/ion and 2.3 atoms/ion, respectively. This results in metal-rich surfaces for both films (Table 2, 4) after Ar$^+$ sputter etching in comparison to *in-situ* measured samples. The thickness of the layer modified by Ar$^+$ sputter-etching can be estimated using Monte-Carlo based TRIM [47] simulations considering the energy of incident ions and their incident angle. With the conditions used in the present study, 1 keV Ar$^+$ ions and 45° incidence angle, for TiN and AlN thin film with ideal bulk density of 5.22 g/cm$^3$ and 3.26 g/cm$^3$, respectively, we estimate the average effective cascade depth defined as the sum of the projected range and straggle of primary recoils (Ti, Al and N). The thickness of the damaged layer is ~2.0 nm for TiN and 2.3 nm for AlN thin films. Therefore, the fraction of the signal intensity contributing to the total intensity of Ti 2p and Al 2p spectra from the Ar$^+$ modified layer, given by eq. $1 - \exp(-d_{ox}/\lambda)$, is ~40 % for TiN and only ~25 % for AlN thin films. Due to the same probing depth of Al $KL_{2,3}L_{2,3}$ Auger emission lines measured by Cr Kα and Al 2p/Al 2s photoelectrons recorded by Al Kα, the fraction of Auger and photoelectron signal intensity contributing to the total intensity of the Auger and photoelectron spectrum from Ar$^+$ modified layer is the same. However, due to a higher sensitivity of the Auger transition to the changes in the local chemical environment it demonstrates a well-defined peak shift and associated change in AP. In contrast, the Ti $L_3M_{2,3}V$ Auger line intensity from the Ar$^+$ modified layer constitutes about ~60 % of the total signal intensity that eventually reflected in the Auger line as a peak shift and significant AP shift.

**Conclusions**

A range of Auger parameters are determined for industrially relevant TiN and AlN thin film by performing combined XPS/HAXPES measurements *in situ*, after ex-posure of the films to ambient air and after subsequent Ar$^+$ sputter-etching. In particular, to the best of our knowledge, α'$_{Ti\ (Ti\ 2p3/2,\ Ti\ KL2,3M4,5)}$ and α'$_{Ti\ (Ti\ 2p3/2,\ Ti\ KL2,3L2,3)}$ for TiN and α'$_{Al\ (Al\ 2p,\ Al\ KL2,3L2,3)}$ and α'$_{Al\ (Al\ 2s,\ Al\ KL2,3L2,3)}$ for AlN thin films are reported for the first time. These parameters provide reliable references for the interpretation of XPS and HAXPES data in future studies on these materials. Our results further reveal that surface oxidation and contamination resulting from air exposure have no measurable impact on the Auger parameters of N, Al, or Ti, regardless of the probing depth of the Auger lines and photoelectrons. Moreover, the minimal influence of air exposure on the intensity of O 1s HAXPES spectra also implies that HAXPES is well suited to study air-transferred AlN thin films. In contrast, Ar$^+$ sputter-etching, commonly employed in the majority of XPS studies, significantly affects the Auger parameters of Al and Ti, and N Auger parameter in AlN thin film. The latter should be carefully considered in future studies of these materials involving the Auger parameter to avoid misinterpretation of the chemical state and bonding environment. The changes in the oxidation state in these materials can be revealed only when avoiding Ar$^+$ sputter-etching or using only bulk-sensitive photoelectron and Auger lines which are insensitive to Ar$^+$ ion-induced surface damage.

25Pshyk *et al.* Empa 2024

**Acknowledgments.** We acknowledge financial support from the Swiss National Science Foundation (R'Equip program, Proposal No. 206021_182987). The authors would like to thank Dr. Giacomo Lorenzin for his assistance and support with sample fabrication and transfer. J.P. and O.P. acknowledge funding from the Swiss National Science Foundation (projects 196980 and 227945).